\documentstyle[sprocl,psfig]{article}

\newcommand{\bra}{\langle}
\newcommand{\ket}{\rangle}

\newcommand{\om}{\omega}
\newcommand{\be}{\begin{equation}}
\newcommand{\ee}{\end{equation}}
\newcommand{\bea}{\begin{eqnarray}}
\newcommand{\eea}{\end{eqnarray}}
\newcommand{\bean}{\begin{eqnarray*}}
\newcommand{\eean}{\end{eqnarray*}}

\begin{document}

\title{NONEQUILIBRIUM FIELDS:\\ EXACT AND TRUNCATED
DYNAMICS\footnote{Talk presented at Strong and Electroweak Matter
(SEWM2000), Marseille, France, 14-17 June, 2000. Based on ref.\ [1], done
in collaboration with Gian Franco Bonini and Christof Wetterich.} 
}

\author{Gert Aarts}

\address{Institut f\"ur theoretische Physik, Universit\"at Heidelberg\\
Philosophenweg 16, 69120 Heidelberg, Germany\\
}

\maketitle 
\abstracts{ 
The nonperturbative real-time evolution of quantum fields out of
equilibrium is often solved using a mean-field or Hartree approximation or
by applying effective action methods. In order to investigate the validity
of these truncations, we implement similar methods in {\em classical}
scalar field theory and compare the approximate dynamics with the full
nonlinear evolution. Numerical results are shown for the early-time
behaviour, the role of approximate fixed points, and thermalization.
}

\section{Testing truncations?}

An understanding of the nonperturbative evolution of quantum fields away
from equilibrium is needed in many physical situations.
Canonical examples are the universe at the end of inflation and
the early stages of a heavy-ion collision. In both cases one would like to
calculate how the energy initially contained in the inflaton or heavy ions
is redistributed over the available degrees of freedom, leading to a hot
universe and a thermal quark-gluon plasma respectively.

Because of the inherent real-time nature and the nonequilibrium setting,
standard euclidean lattice methods are not applicable and approximations
have to be introduced. In the last ten years or so a lot of attention has
been given to mean-field approximations, in which the dynamics of a mean
field coupled to Gaussian (quadratic) fluctuations is solved
self-consistently.  Well-known examples are the Hartree and leading-order
large $N_f$ approximations, with $N_f$ the number of scalar or fermion
fields.  Several paths going beyond homogeneous mean fields have been
explored recently as well. For a list of references, see [1]. It goes
without saying that the use of approximations will introduce deviations
from the true evolution in quantum field theory.  In order to gain trust
in a particular method, it would be helpful to compare the approximate
dynamics with the exact one.
Unfortunately, in QFT such tests are not easy since the exact
nonperturbative evolution is not known. Therefore we decided to use
classical fields, regularized on a lattice, instead. In many aspects
classical fields are similar to quantum fields, e.g.\ scattering is
present and the role of the thermodynamic limit at fixed lattice spacing
can be investigated. Furthermore, most approximation methods can be
implemented as well, as they do not involve $\hbar$ directly. Last but not
least, the `exact' evolution can be calculated numerically which allows
for a direct comparison.

\section{Classical Hartree and beyond}

We consider classical $\phi^4$ theory in $1+1$ dimensions. The dynamics is
determined by the equation of motion $\partial_t^2\phi(x,t) =
(\partial^2_x-m^2)\phi(x,t)-\lambda \phi^3(x,t)/2$, given initial
conditions $\phi(x,0)$ and $\pi(x,0)=\partial_t\phi(x,0)$ that are taken
from a probability distribution $\rho[\pi,\phi]$. Ensemble averages are
denoted with $\bra\cdot\ket$.  The exact nonlinear evolution can be
calculated by sampling initial conditions from $\rho$ and solving the
e.o.m.\ numerically for each of them.

A classical Hartree-type of approximation can be implemented by replacing
$\lambda \phi^3$ with $3\lambda \phi\bra\phi^2\ket$ in the equation above.
For translationally invariant systems
$\bra\phi^2\ket \equiv \bra\phi^2(x,t)\ket$ is independent of $x$ and the
Hartree e.o.m.\ can be written conveniently in momentum space as
$\partial_t^2\phi(q,t) =
-\bar\om_q^2\phi(q,t)$. The effective frequency squared is
$\bar\om_q^2=\om_q^2+\frac{3}{2}\lambda \bra\phi^2\ket$, with
$\om_q^2=q^2+m^2$. The unequal-time two-point function $S(x-y;t,t') = \bra
\phi(x,t)\phi(y,t')\ket$ obeys in this approximation the usual mean-field
equation of motion $(\partial_t^2+\bar\om_q^2)S(q;t,t')=0$.
For the Hartree approximation with inhomogeneous mean fields in the
quantized version of this model, see [2].

Beyond the Hartree approximation unequal-time formulations typically
become nonlocal in time. This is a disadvantage when numerical solutions
are required. Therefore we reformulate the Hartree dynamics in terms of
three basic equal-time two-point functions: $G_{\psi\psi'}(x-y,t) =
\bra\psi(x,t)\psi'(y,t)\ket$, with $\psi=\{\pi,\phi\}$. In terms of these
the Hartree evolution equations read
\bea
\nonumber
&&\partial_t G_{\phi\phi}(q,t) = 2G_{\pi\phi}(q,t),\\
\label{eqLO}
&&\partial_t G_{\pi\phi}(q,t) = -\bar\om_q^2 G_{\phi\phi}(q,t) +
G_{\pi\pi}(q,t),\\
\nonumber
&&\partial_t G_{\pi\pi}(q,t) = -2\bar\om_q^2G_{\pi\phi}(q,t).
\eea
These equations conserve $\alpha^{-2}(q) =
G_{\phi\phi}(q,t)G_{\pi\pi}(q,t) - G^2_{\pi\phi}(q,t)$ for each $q$.

The evolution equations beyond Hartree include dynamical equal-time
four-point functions. In the case of an $O(N_f)$ theory they contain all
$1/N_f$ corrections. Details can be found in [1]. 
It remains to specify the
initial probability distribution. We take a Gaussian ensemble with $ \bra
\psi(q,0)\psi'(q',0)\ket =
G_{\psi\psi'}(q,0)2\pi\delta(q+q')\delta_{\psi\psi'}$, and
\be
\label{eqens}
G_{\phi\phi}(q,0) = T_0/(q^2+m^2), \;\;\;\;\;\;\;\;
G_{\pi\pi}(q,0)= T_0.
\ee
Note that this corresponds to the equilibrium ensemble when $\lambda=0$.

\begin{figure}
\centerline{
\psfig{figure=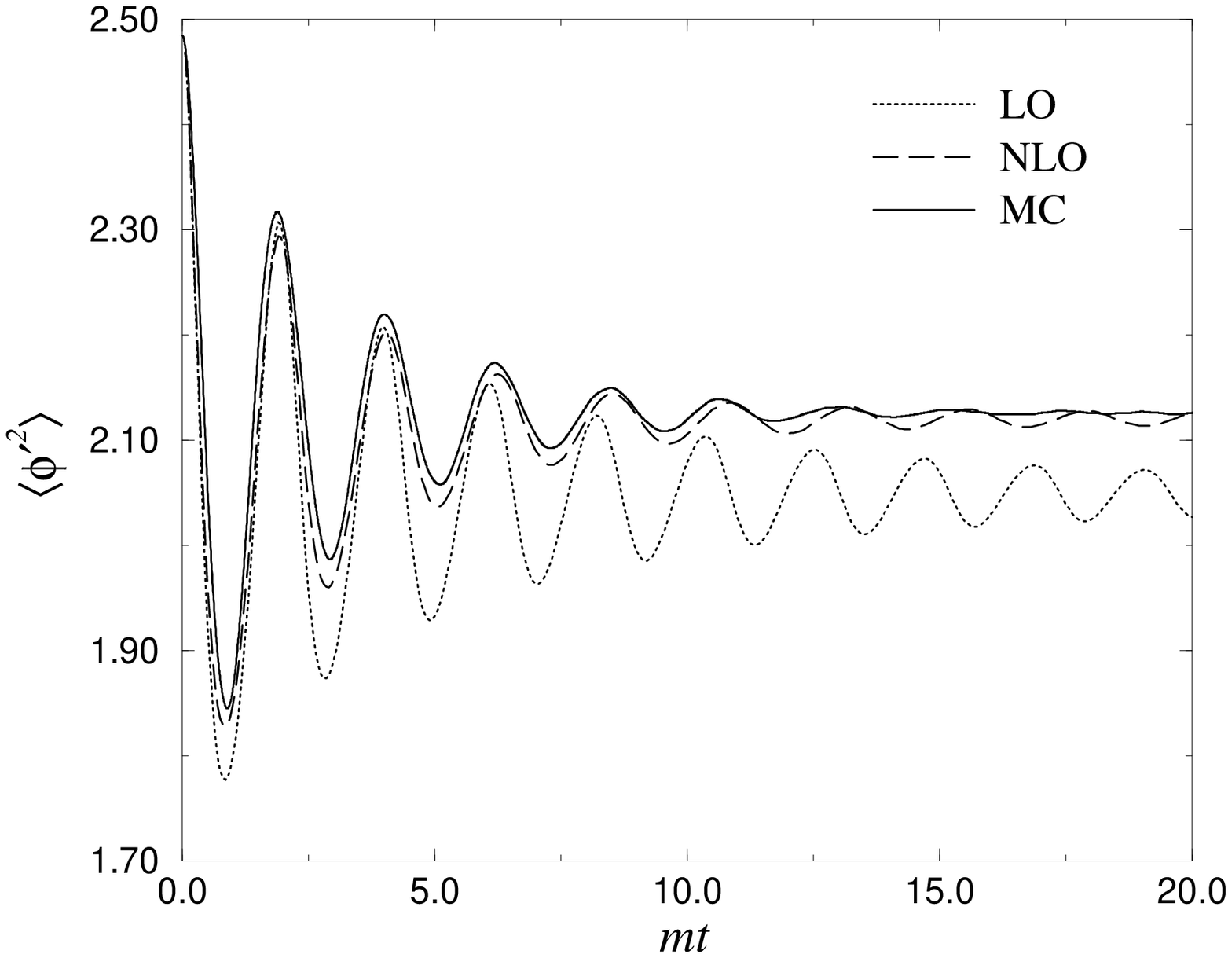,height=4.9cm}
\hspace{-0.3cm}
\psfig{figure=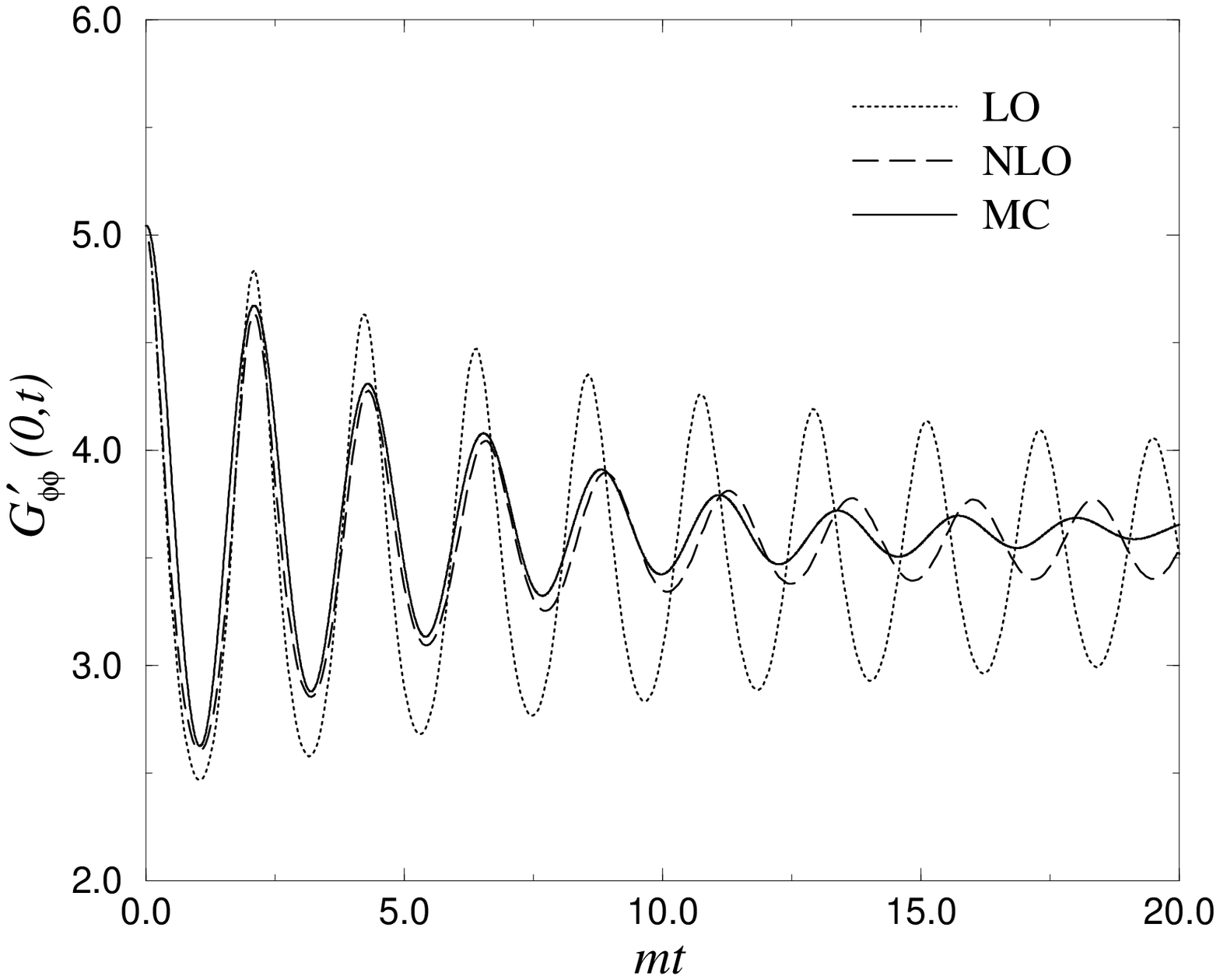,height=4.9cm}
}
\vspace{-0.5cm}
\caption{Early-time evolution in the Hartree approximation (LO),
with the inclusion of four-point functions (NLO), and `exact'
(MC). Left: mean
field squared $\bra \phi'^2\ket$. Right: zero-mode 
$G_{\phi\phi}'(q=0,t)$. The initial temperature of the Gaussian
ensemble is $T'_0=5$.
}
\label{figearly}
\end{figure}

\section{Early time, fixed points, and thermalization}

Typical behaviour at early times is shown in Fig.\ \ref{figearly} for two
observables. We see that the LO evolution is in qualitative agreement with
MC result. At NLO the agreement is impressive. For numerical
details I refer again to [1]. Note that primes indicate dimensionless
quantities and $T_0'\equiv 3\lambda T_0/m^3$.
	
\begin{figure}
\centerline{
\psfig{figure=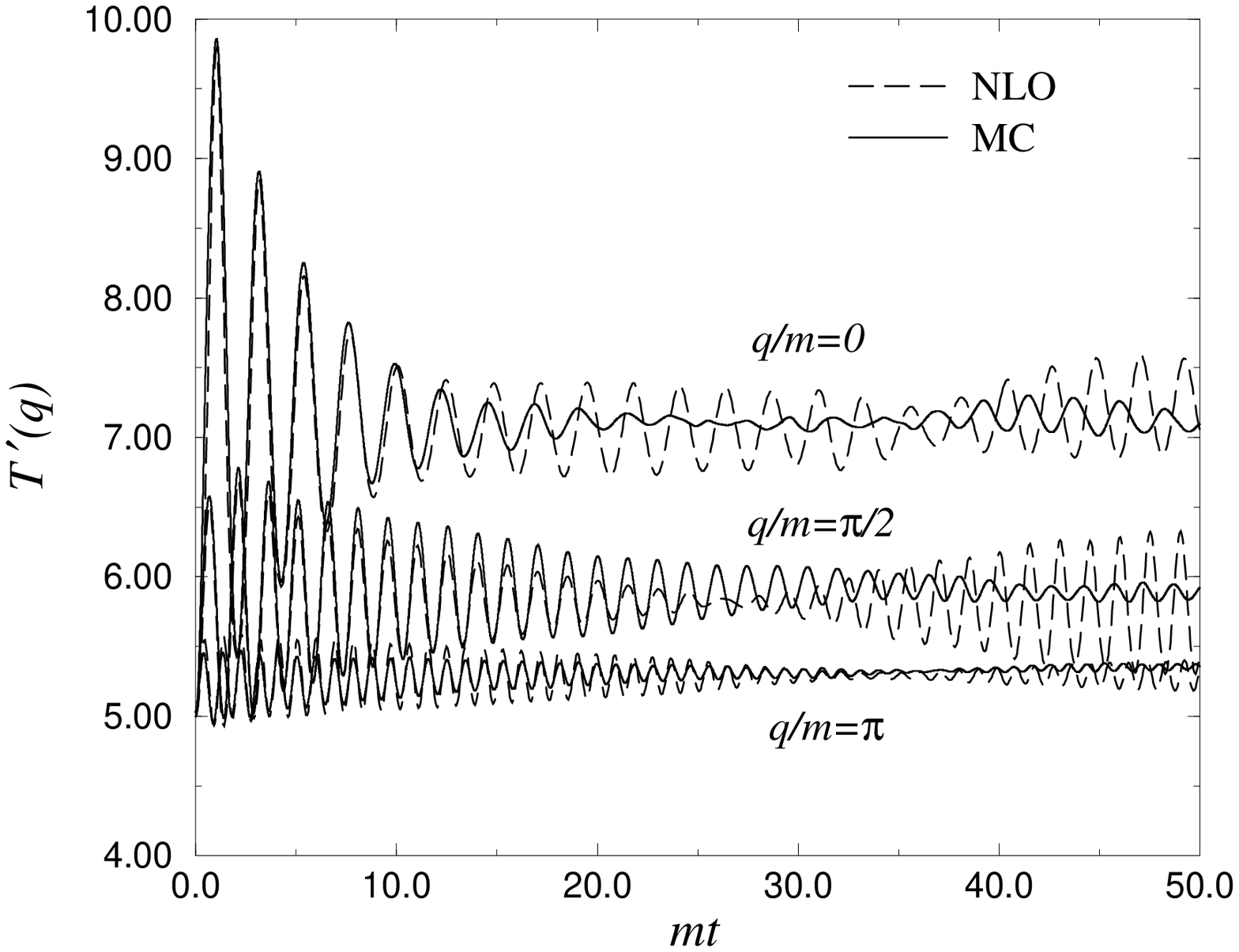,height=4.9cm}
\hspace{-0.3cm}
\psfig{figure=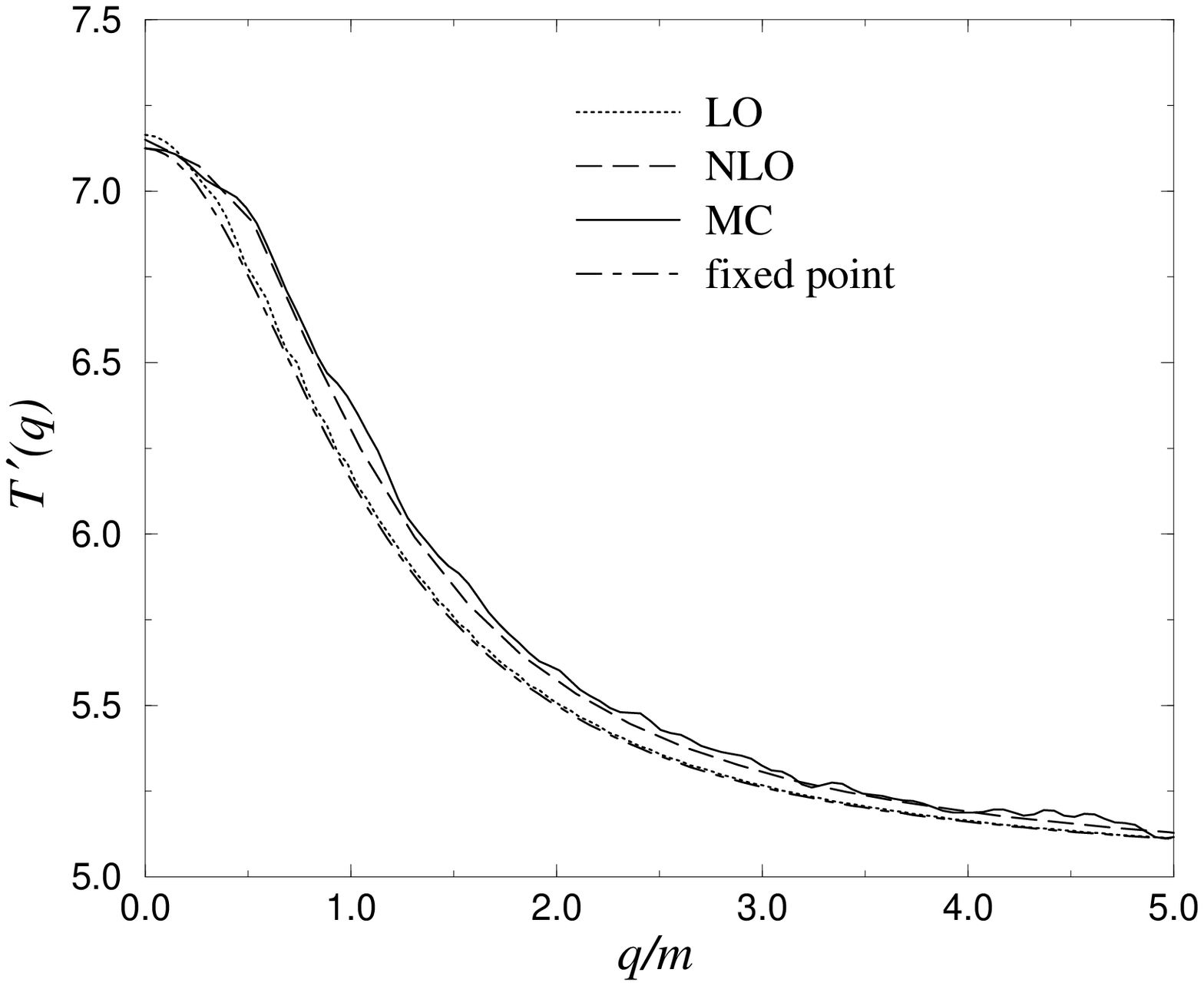,height=4.9cm}
}
\vspace{-0.5cm}
\caption{Nonthermal behaviour. Left: effective temperature $T'(q,t)$
for 3 momentum modes. Right: time-averaged temperature profile between 
$0<mt<50$, the fourth line is the analytic expression (\ref{eqprof}) at
the fixed point in the Hartree approximation ($T_0'=5$).
}
\label{figprofile}
\end{figure}

An important issue in nonequilibrium field theory is the approach to
equilibrium. One of the crucial tests for a truncated evolution is whether
it is able to describe the thermalization regime or if the approximation
breaks down before. In a classical field theory, a convenient two-point
function to investigate is $G_{\pi\pi}(q,t)\equiv T(q,t)$. Using the
equilibrium partition function as a guide, we see that this correlator can
be interpreted as the effective temperature for a momentum mode $q$. In
equilibrium, $T(q) = T$ for all $q$. In Fig.\ \ref{figprofile} (left) we
show the time evolution of $T(q,t)$ for three momentum modes (NLO, MC
only). It is clear
that the modes have a different effective temperature. The system seems
to be quasi-stationary but is not thermal. 

A remarkable feature is that this behaviour can be understood completely
in terms of fixed points in the truncated evolution equations. The
stationary points in the Hartree approximation are easily found from Eq.\
(\ref{eqLO}), and obey
\be
\label{eqfp}
G^*_{\pi\pi}(q) =  \bar\om_q^{*2} G^*_{\phi\phi}(q), \;\;\;\;\;\;\;\;
G^*_{\pi\phi}(q) = 0,
\ee
with $\bar\om_q^{*2}=\om_q^2+\frac{3}{2}\lambda\bra\phi^2\ket^*$. These
relations can be completed with the expression for the conserved
combination at the fixed point
\be
\label{eqalphafp}
\alpha^{-2}(q) = T_0^2/\om_q^2 = G^*_{\pi\pi}(q)G^*_{\phi\phi}(q). 
\ee
The first equality follows from the initial ensemble (\ref{eqens}).
Eqs.\ (\ref{eqfp}) and (\ref{eqalphafp}) specify the fixed point
completely. We find 
\be
G^*_{\phi\phi}(q) = \frac{T_0}{\bar\om^*_q\om_q}, \;\;\;\;\;\;\;\;
\bra \phi^2\ket^* = \int \frac{dq}{2\pi}\,
\frac{T_0}{\bar\om^*_q\om_q},
\ee
(the second expression is a gap equation for $\bra\phi^2\ket^*$ and can
be written in terms of elliptic functions)
and 
\be
\label{eqprof}
G^*_{\pi\pi}(q) = T_0\frac{\bar\om^*_q}{\om_q} =
T_0\left[ 1 +
\frac{3}{2}\frac{\lambda\bra\phi^2\ket^*}{\om_q^2}\right]^{1/2},
\ee
i.e., a nonthermal profile of $G^*_{\pi\pi}(q)$ at the fixed point. In
order to see whether this fixed-point profile is realized in the dynamical
evolution, we show the time average of $G_{\pi\pi}(q,t)$ for the Hartree,
NLO, and full evolution in Fig.\ \ref{figprofile} (right). It is clear
that the analytic expression not only describes the Hartree result, but
also the profile from the NLO and in particular the full nonlinear
evolution surprisingly well. This can be interpreted as a justification
for the use of a Hartree approximation in this time interval.

\begin{figure}
\centerline{
\psfig{figure=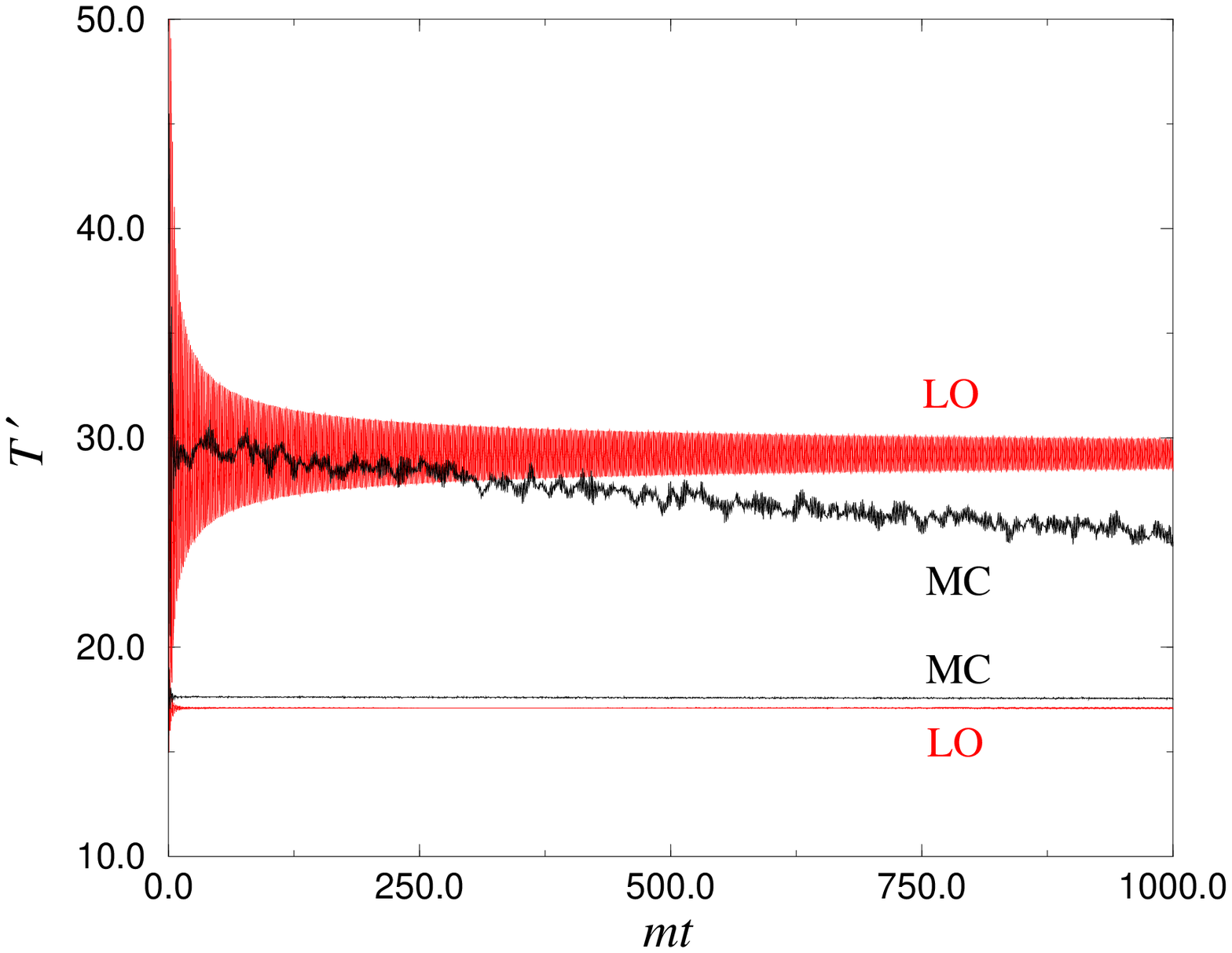,height=4.9cm}
\hspace{-0.3cm}
\psfig{figure=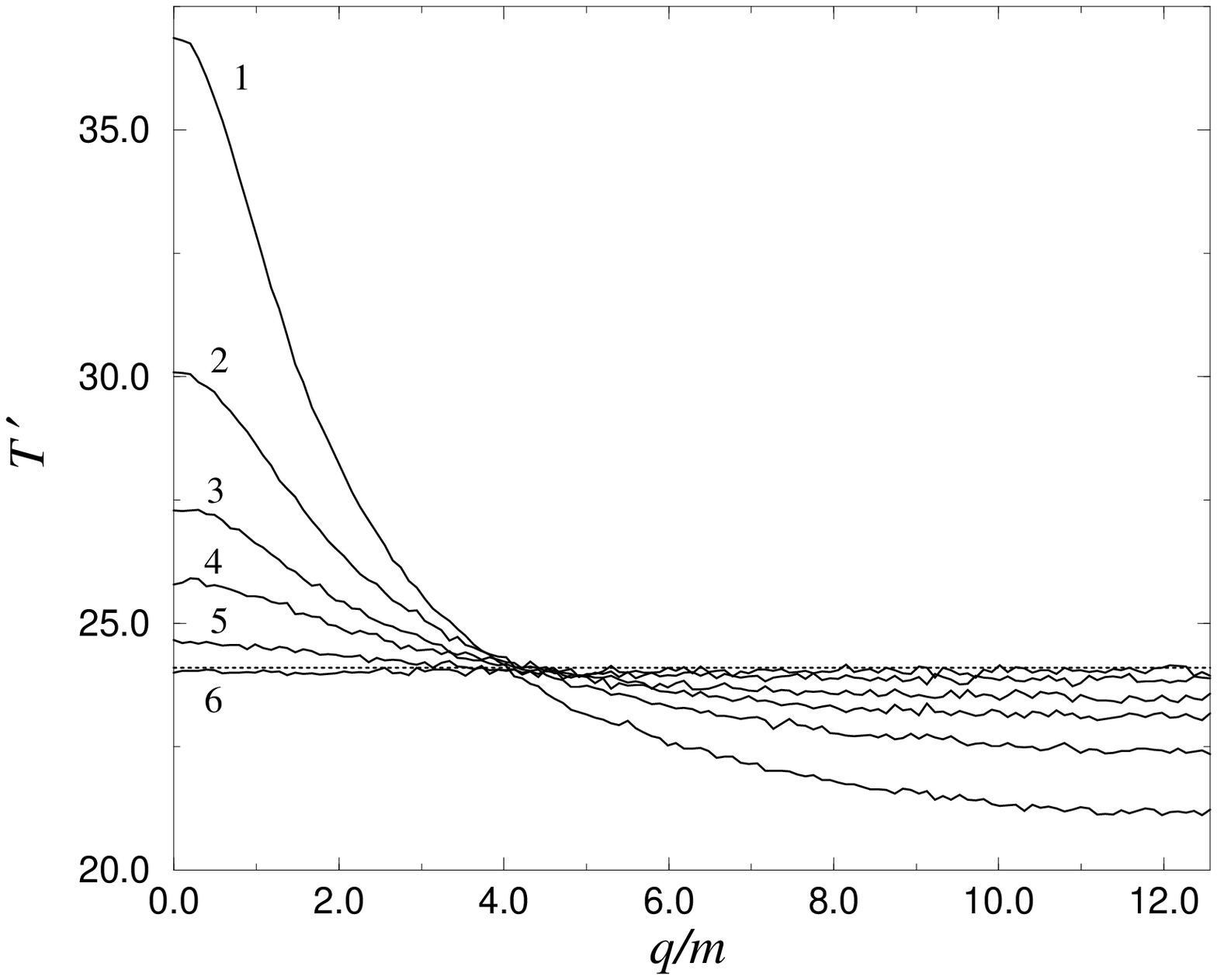,height=4.8cm}
}
\vspace{-0.5cm}
\caption{Fate of the fixed point. Left: zero-mode temperature
$T'(q=0,t)$ for LO and MC ($T_0'=15$). In the Hartree approximation the
fixed point governs the dynamics for late times as well, whereas the full
nonlinear dynamics shows thermalization. 
The average temperatures $T'(t)$ appear as straight lines. 
Right: time-averaged snapshots of the temperature profile (MC only,
$T_0'=20$, line 1: $0<mt<1500$, $\ldots$, line 6: $13500<mt<15000$). 
The fixed-point shape slowly disappears and the spectrum becomes thermal
(flat). 
}
\label{figspec}
\end{figure}

The fate of the fixed point can be determined by being patient. In Fig.\
\ref{figspec} (left) we show the evolution of $T(q=0,t)$, for a higher
initial temperature $T_0$ as before. Also shown are the average
temperatures $T(t)=N^{-1}\sum_q T(q,t)$. The Hartree evolution remains to
be controlled by its fixed point. In the full nonlinear evolution on
the other hand, all momentum modes obtain the same temperature. The change
from a fixed-point profile to a thermal profile is shown in Fig.\
\ref{figspec} (right) for the full evolution only. Due to instabilities, 
it was not possible to reach such late times using the NLO evolution [1].

\section{Summary}

In order to gain insight in how well truncated dynamics reproduces the
full evolution in nonequilibrium quantum field theory, we studied the
corresponding classical problem.  We stress that the aim was not to use
classical fields to approximate the quantum theory, but rather to study
truncated dynamics in a field theory where the exact evolution is
available. Our findings are that while at early times the truncated
evolution works remarkably well and the outcome can be characterized
nicely in terms of fixed points of the associated Hartree equations, it
breaks down before the thermalization regime is reached.

\section*{Acknowledgments}

\noindent
I would like to thank Jan Smit and Jeroen Vink for discussions. 
This work is supported by the TMR network FMRX-CT97-0122.

\end{document}